\documentclass[twocolumn,showpacs,prb]{revtex4-1}
\usepackage{bm}
\usepackage{graphicx}
\usepackage{epsfig}
\usepackage{amsmath}
\usepackage{color}
\usepackage{amssymb}
\usepackage{array}
\usepackage{longtable}
\usepackage[colorlinks = true]{hyperref}
\usepackage{epstopdf}
\usepackage{tabularx}

\numberwithin{equation}{section}

\begin{document}

\title{The Stefan outflow in a multicomponent vapor-gas atmosphere around a  droplet and its role for cloud expansion}

\author{A.\,E.\,Kuchma}
\author{A.\,K.\,Shchekin}
\author{D.\,S.\,Martyukova}

\affiliation{Department of Statistical Physics, Faculty of Physics, St. Petersburg State University, 7/9 Universitetskaya nab., St. Petersburg, 199034, Russia}

%************************************************
\begin{abstract}
%************************************************

A new comprehensive analysis of Stefan's flow caused by a free growing droplet in vapor-gas atmosphere with several condensing components is presented. This analysis, based on the nonstationary heat and material balance and  diffusion transport equations, shows the appearance of the Stefan inflow  in the vicinity of the growing droplet and the outflow at large distances from the droplet as a consequence of nonisothermal condensation. For an ensemble of droplets in the atmospheric cloud, this flow provides an increase of the total volume of the cloud, which can be treated as cloud thermal expansion and leads to floating the cloud as a whole due to buoyancy. We have formulated the self-similar solutions of the nonstationary diffusion and heat conduction equations for a growing multicomponent droplet and have derived analytical expressions for the nonstationary velocity profile of Stefan's flow and the expansion volume of the vapor-gas mixture around the growing droplet. To illustrate the approach, we computed these quantities in the case of  droplet of stationary composition in air with several specific vapors (${\mathrm{C_2H_5OH/H_2O}}$; ${\mathrm{H_2SO_4/H_2O}}$; ${\mathrm{H_2O}}$).
\end{abstract}

\maketitle
%************************************************
\section*{Introduction}\label{sec:intro}
%************************************************

Growth and evaporation of small droplets are key phenomena in the physics of aerosols and clouds. The regularities of phase changes and latent heat release have a global influence on the Earth atmosphere and climate (\cite{Kessler1969}, \cite{Heintzenberg2009}, \cite{Pruppacher2010}). Cloud microphysical schemes are a central part of every model of the atmosphere. In numerical weather simulation, they are important for quantitative precipitation forecasts.

Our goal in this paper is to give a detailed analysis of one of mechanisms for thermal cloud expansion in the atmosphere. This mechanism is associated with the Stefan flow caused by nonisothermal growth of small droplets in the cloud. The Stefan flow of the vapor-gas mixture around a droplet compensates the diffusive flux of noncondensable gas molecules from the impenetrable surface of the growing droplet. This flow under ordinary atmospheric conditions is usually considered to be a small effect (\cite{Fuchs1959a, Clement2009a, Kuchma2011a, Kuchma2012a}). However there are some atmospheric situations where the Stefan flow may be important. The particle scavenging efficiency of vapor-grown ice crystals falling from mixed clouds proves to be very high due to the Stefan flow which pushes airborne particles away from the surface of the supercooled droplets evaporating in the vicinity of an ice crystal (\cite{Vittori1984}). The influence of the Stefan flow on the heat balance of a char particle should be taken into account in oxy-fuel combustion (temperature ${\approx 1400}$ K), because, the presence of Stefan's flow greatly weakens the heat transfer by conduction and accelerates the heat loss from the particle, which makes the particle temperature lower (\cite{Zhou2013}).  Another effect, as follows from \cite{Kuchma2016a}, owes to nonstationary heating a vapor-gas mixture around a growing droplet (due to releasing the condensation heat) which produces the Stefan outflow at large distances from the droplet. For an ensemble of droplets in the cloud, such effect results in increase of the total volume of the cloud that can be treated in physics of clouds as cloud thermal expansion followed by the rise of the cloud as a whole due to buoyancy. In the case of several vapors, also the vertical structure of a cloud with droplets of different size and composition can change. Thus we have an example where a small effect on a level of a single droplet becomes an origin for strong mass and heat transport on the macroscopic scale of a cloud.

We will formulate below the theory of the Stefan flow in the case of multi-component nonisothermal and nonstationary growth of a free droplet with non-ideal composition. As a first step, we will derive in Sec.~\ref{sec:flow} the general expressions for the velocity of the Stefan flow based on the material and enthalpy balance equations at nonstationary diffusion. In Sec.~\ref{sec:self-similar}, we will consider the self-similar solutions of the nonstationary diffusion and heat conduction equations for a growing multicomponent droplet and obtain analytical expressions for the whole nonstationary velocity profile of Stefan's flow and the expansion volume of the vapor-gas mixture around the growing droplet. Finally in Sec.~\ref{sec:simulating}, we will illustrate the approach by computing velocity profile of Stefan's flow and the expansion volume of the vapor-gas mixture around the growing droplet for droplets of stationary composition in air with several specific vapors (binary mixture of ethanol-water ${\mathrm{C_2H_5OH/H_2O}}$ vapors; binary mixture of sulfuric acid-water ${\mathrm{H_2SO_4/H_2O}}$ vapors, and pure water vapor ${\mathrm{H_2O}}$).

%%%%%%%%%%%%%%%%%%%%%%%%%%%%%%%%%%%%%%%%%%%%%%%%%%%%%%%%%%%%%%%%
%%%%%%%%%%%%%%%%%%%%%%%%%%%%%%%%%%%%%%%%%%%%%%%%%%%%%%%%%%%%%%%%

\section{General regularities for a multicomponent Stefan flow}\label{sec:flow}

Consider a small free spherical droplet suspended in the ideal mixture of several condensable vapors and a non-condensable carrier gas. The condensable components are assumed to be miscible in the droplet at any ratios, and the solution in the droplet may be non-ideal. The droplet is tiny enough to provide that the specific times of establishing internal thermal and chemical equilibrium in the droplet are small in comparison with the specific time of changing the droplet radius in time. As a consequence, the droplet is internally uniform in composition and temperature. The droplet can grow or evaporate depending on the ratio between the actual concentrations of vapors far away from the droplet and the concentrations of the vapors at saturation with the solution in the droplet. We will take into account releasing or absorbing the latent heat at condensation or evaporation. As a result, the temperatures of the droplet and surrounding vapor-gas mixture may differ. The regimes of vapor and heat transfer to or from the droplet are diffusive. The total pressure in the vapor-gas mixture is considered to be constant. Gravity is neglected.

We set the origin of the coordinate system in the center of the spherical droplet with radius ${R(t)}$ changing in time ${t}$. In view of the spherical symmetry, the local vapor concentration ${n_i(r,t)}$ of the condensable component ${i}$  ${(i = 1,2 ,\dots,k)}$, i.e., the local number of molecules of ${i}$th component per unit volume, and the local temperature ${T(r,t)}$ in the vapor-gas mixture at any time moment ${t}$ depend only on distance ${r}$ from the center of the droplet to the observation point. The corresponding concentration of the non-condensable carrier gas will be denoted by ${n_g(r,t)}$. Thereby the total local volume concentration of molecules in the vapor-gas mixture is determined as
\begin{equation}
\label{total amount_n}
   n(r,t) = n_g(r,t) + \sum_{i=1}^k n_i(r,t),
\end{equation}
The boundary conditions for the vapor concentrations ${n_i(r,t)}$ and temperature ${T(r,t)}$ can be written in the form
\begin{equation}
\label{cond_ni}
\begin{gathered}
     n_i(r,t)\xrightarrow[r\to \infty]{} n_{i0}, \\
      n_i(r,t)\Bigl |_{r=R(t)}= n_{i\infty}(\{x\},T_d), \quad  i=1,2\dots k,
\end{gathered}
\end{equation}
\begin{equation}
\label{cond_n}
     n(r,t)\xrightarrow[r\to \infty]{} n_{0} = n_{g0} + \sum_{i=1}^k n_{i0},
\end{equation}
\begin{equation}
\label{cond_T}
     T(r,t)\xrightarrow[r\to \infty]{} T_0, \quad T(r,t)\Bigl |_{r=R(t)}= T_d,
\end{equation}
where ${n_{i0}}$ and ${n_{g0}}$ are the bulk concentration of ${i}$th component and carrier gas, respectively, $T_0$ is the temperature far away from droplet, ${n_{i\infty}(\{x\},T_d)}$ is the equilibrium concentration of ${i}$th component over a planar surface of the liquid solution at droplet temperature ${T_d}$ (we neglect the effect of the curvature of droplet surface on the ${i}$th vapor equilibrium pressure and concentration) and composition ${\{x\}} \equiv \{x_1, x_2, \dots x_k\}$. The molar concentration ${x_i}$ of condensible components in the droplet (the total number of droplet components coincides with the number of the vapor components in the vapor-gas mixture) is defined by
\begin{equation}
\label{xx}
    x_i(t) = \frac{N_i(t)}{N(t)}, \quad i=1,2\dots k.
\end{equation}
Here ${N_i}$ is the number of molecules of ${i}$th component in the droplet, and ${N \equiv \sum \limits_{i=1}^k N_i}$ is the total number of molecules in the droplet.

\subsection{Velocity of a multicomponent Stefan flow}\label{sec:flow_near}
The total number of all molecules inside an imaginary sphere around droplet with radius  ${R_1(t)}$, which expands with the velocity ${dR_1(t)/dt =u(r=R_1(t),t)}$ of the hydrodynamic molecular flow of the vapor–gas mixture (i.e., the velocity of the Stefan flow), does not depend on time and is distributed between the droplet and the vapor-gas phase as
{\sloppy

}
\begin{equation}
\label{total amount}
   N(t)+4\pi \int\limits_{R(t)}^{R_1(t)} dr r^2 n(r,t)=\mathrm{const},
\end{equation}
Differentiating both sides of  eq.\eqref{total amount} with respect to time and setting ${dN(t)/dt \equiv \dot{N}(t)}$,  ${dR(t)/dt \equiv \dot{R}(t)}$, and ${R_1(t)=r}$ allows us to express the Stefan velocity  ${u(r,t)}$ in the form
{\sloppy

}
\begin{equation}
\label{u_full}
\begin{gathered}
  u(r,t)=-\frac{\dot{N}(t)}{4 \pi r^2 n(r,t)} + \frac{R^2(t) \dot{R}(t)}{r^2}\frac{n(R(t),t)}{n(r,t)} - \\
   - \frac{1}{r^2 n(r,t)}\int\limits_{R(t)}^{r} dr_1 r_1^2 \frac{\partial n(r,t)}{\partial t}.
\end{gathered}
\end{equation}

At diffusive regime of droplet growth, the droplet composition ${\{x\}}$ and the droplet temperature ${T_d}$ become independent of time after some transient stage (\cite{Kulmala1993a}, \cite{Kuchma2011b, Kuchma2011c},\cite{Kuchma2013a}). Because the time of this transient stage is small, we will neglect this stage and consider that the droplet composition ${\{x\}}$ and the temperature of droplet ${T_d}$ do not depend on time, i.e. ${dx_i/dt=0}$ (${i=1,2\dots k}$) and ${dT_d/dt=0}$, respectively. Recognizing that the pressure is constant in the ideal vapor-gas mixture, we have the following relations between total local volume concentration of molecules and local temperature in the vapor-gas mixture
{\sloppy

}
\begin{equation}
\label{ideal}
     n(r,t)T(r,t) = n_0 T_0,
\end{equation}
\begin{equation}
\label{d_ideal_n}
     \frac{\partial n(r,t)}{\partial t} = - \frac {n_0 T_0}{T^2(r,t)}\frac{\partial T(r,t)}{\partial t}.
\end{equation}
Using eqs.\eqref{ideal} and \eqref{d_ideal_n} and the relationship for the droplet volume ${V_d}$,
\begin{equation}
\label{vol}
    V_d(t)=\displaystyle \frac{4\pi}{3}R^3=\sum_{i=1}^k v_{il}(\{x\},T_d)x_i N \equiv v_{l}(\{x\},T_d) N,
\end{equation}
where ${v_{il}(\{x\},T_d)}$ is the partial volume per molecule of ${i}$th component and ${v_{l}(\{x\},T_d)}$ is the mean molecular volume in the droplet with total composition ${\{x\}}$ and temperature ${T_d}$, allows us to express the velocity of the Stefan flow from \eqref{u_full} as
\begin{equation}
\label{u_full_v}
\begin{gathered}
  u(r,t)=\dot{R}\frac{R^2}{r^2}\frac{T(r,t)}{T_d}\Bigg( 1-\frac{T_d}{v_{l}(\{x\},T_d)
  n_0T_0}+ \\
  + \frac{T_d}{R^2\dot{R}}\int\limits_{R(t)}^{r} dr_1 r_1^2 \frac{1}{T^2(r_1,t)}\frac{\partial T(r_1,t)}{\partial t} \Bigg).
\end{gathered}
\end{equation}
Here we took into account that the composition ${\{x\}}$ and the droplet temperature ${T_d}$ are fixed in time. Eq. \eqref{u_full_v} is the general expression for the velocity of the Stefan flow in the case of nonisothermal multicomponent growth of a droplet with non-ideal solution at stationary composition and temperature.

It strictly follows from eq.\eqref{u_full_v} at ${\dot{R}>0}$ and ${n_0 v_l \ll 1}$ that
\begin{equation}
\label{dop_6}
  u(r,t)\xrightarrow[r\to R(t)]{} -\dot{R}\frac{T_d}{v_{l}(\{x\},T_d)n_0T_0} <0.
\end{equation}

Thereby, the Stefan flow is directed to the droplet in the droplet vicinity. As is seen from eq.\eqref{u_full_v}  at ${n_0 v_l \ll 1}$ and at isothermal condensation or stationary heat conduction when ${\partial T(r,t)/ \partial t =0}$, the velocity of the Stefan flow is directed to the growing droplet at any  ${r \ge R(t) }$
\begin{equation}
  u(r,t)  = -\dot{R}\frac{R^2}{r^2}\frac{T(r,t)}{v_{l}(\{x\},T_0)n_0 T_0} <0.
\end{equation}\\

%%%%%%%%%%%%%%%%%%%%%%%%%%%%%%%%%%%%%%%%%%%%%%%%%%%%%%%%%%%%%%%%
%%%%%%%%%%%%%%%%%%%%%%%%%%%%%%%%%%%%%%%%%%%%%%%%%%%%%%%%%%%%%%%%

\subsection{The hydrodynamic flow rate far from the droplet}\label{sec:flow_far}

In the case of non-stationary heat conduction, the integral term in brackets in eq.\eqref{u_full_v} is able to change the sign of the velocity ${u(r,t)}$ at large ${r}$. To check its role, we consider below the balance of the total enthalpy of the system consisting of droplet and vapor-gas mixture. Let us express the total enthalpy ${H}$ of the droplet and surrounding vapor-gas atmosphere within the sphere of radius ${R_2(t)}$ as
\begin{equation}
\label{law of conservation of enthalpy}
\begin{gathered}
  H = h_{l}(\{x\},T_d)N+ \\
  + 4\pi \int\limits_{R(t)}^{R_2(t)} dr r^2 \Bigg( h_g(T(r,t))n_g(r,t)+
   \sum_{i=1}^k h_{i}(T(r,t))n_i(r,t)\Bigg),
\end{gathered}
\end{equation}
where
\begin{equation}
\label{heat_h}
  h_{l}(\{x\},T_d) \equiv \sum_{i=1}^k h_{il}(\{x\},T_d)x_i
\end{equation}
is the mean enthalpy per molecule in the droplet solution, ${h_{il}(\{x\},T_d)}$ is the partial enthalpy of ${i}$th component in droplet solution, ${h_{g}(T(r,t))}$ and ${h_{i}(T(r,t))}$ are the enthalpies per molecule  of carrier gas and of ${i}$th component of vapor, respectively. If we take the radius ${R_2}$ as the radius of the sphere which is located outside the diffusion layer (this is possible only for nonstationary diffusion and heat conduction) and which boundary shifts with the velocity ${u(r,t)}$ of the hydrodynamic molecular flow of the vapor-gas mixture, i.e. ${dR_2(t)/dt=u(r=R_2(t),t)}$, the enthalpy ${H}$ will conserve in time.
{\sloppy

}

Differentiating both sides of eq.\eqref{law of conservation of enthalpy} with respect to time at fixed ${\{x\}}$ and ${T_d}$ (we recall that the transition stage of the droplet growth is considered to be fast comparatively with the time of observation) and setting ${R_2(t)=r}$ allow us to express the Stefan velocity ${u(r,t)}$ outside the diffusion shell around the droplet in the form
\begin{equation}
\label{prom3}
\begin{gathered}
  4 \pi \left(h_g(T_0)n_{g0} + \sum_{i=1}^k h_i(T_0)n_{i0}\right) r^2 u(r,t)=
  \\= -h_l(\{x\}, T_d) \dot{N} +
  + 4 \pi R^2(t) \dot{R}(t) 
  \Bigg(h_g(T_d)n_g(R(t),t) + \\ +\sum_{i=1}^k h_i(T_d)n_i(R(t),t)\Bigg) -\\ - 4 \pi  \int\limits_{R(t)}^{r} dr_1 r_1^2 \frac{\partial}{\partial t}\Bigg(h_g(T(r_1,t))n_g(r_1,t) + \\
   + \sum_{i=1}^k h_i(T(r_1,t))n_i(r_1,t)\Bigg).
\end{gathered}
\end{equation}
With the help of eq.\eqref{total amount_n}, it is convenient to write
\begin{equation}
\label{h_g}
\begin{gathered}
  h_g(T(r,t))n_g(r,t) + \sum_{i=1}^k h_i(T(r,t))n_i(r,t) =\\
  = h_g(T(r,t))n(r,t) + \sum_{i=1}^k \left(h_i(T(r,t)) - h_g(T(r,t))\right)n_i(r,t).
\end{gathered}
\end{equation}

In view of eq.\eqref{ideal} we have
\begin{equation}
\label{dop1}
    h_g (T(r,t))n(r,t) = c_{g} T(r,t) n(r,t) = c_{g} T_0 n_0,
\end{equation}
where ${c_{g}}$ is the heat capacity at constant pressure per carrier gas molecule.
Substituting eq.\eqref{dop1} into eq.\eqref{h_g} and neglecting the dependence of ${c_{g}}$ on temperature and differentiating with respect to time yields
\begin{equation}
\label{promm3}
\begin{gathered}
    \frac{\partial}{\partial t} \left[ h_g (T(r,t))n_g(r,t) + \sum_{i=1}^k h_i(T(r,t)) n_i(r,t) \right] = \\
    = \frac{\partial}{\partial t} \sum_{i=1}^k \left( h_{i} (T(r,t)) -  h_g(T(r,t)) \right) n_i(r,t).
\end{gathered}
\end{equation}

Let us first consider the particular case when ${h_i = h_g}$, ${i=1,2\dots k}$, i.e. the enthalpies per molecule of carrier gas and all vapor components are equal. As follows from eqs.\eqref{promm3} and \eqref{cond_n}, eq.\eqref{prom3} can be rewritten in this particular case as
\begin{equation}
\label{dop2}
\begin{gathered}
    h_g (T_0) n_0 r^2 u(r,t) = - R^2(t) \dot{R}(t) h_l(\{x\},T_d) v_l^{-1}(\{x\},T_d) + \\
    + R^2(t) \dot{R}(t) \left[ h_g(T_d) n_g(R(t),t) + \sum_{i=1}^k h_i(T_d) n_i(R(t),t) \right] ,
\end{gathered}
\end{equation}
where we used eqs.\eqref{vol}. Introducing the mean evaporation heat ${q(\{x\}, T_d)}$ per molecule in the liquid with composition ${\{x\}}$ and temperature ${T_d}$  as
\begin{equation}
\label{q}
\begin{gathered}
  q(\{x\},T_d) \equiv \sum_{i=1}^k \left[ h_{i}(T_d)-h_{il}(\{x\},T_d) \right] x_i = \\ =\sum_{i=1}^k q_{il}(\{x\},T_d) x_i ,
\end{gathered}
\end{equation}
where ${q_{il}(\{x\},T_d)}$ is the partial evaporation heat of $i$th component, we find from eq.\eqref{dop2}
\begin{equation}
\label{dop3}
\begin{gathered}
    h_g (T_0) n_0 r^2 u(r,t) = R^2(t) \dot{R}(t) q(\{x\},T_d) v_l^{-1}(\{x\},T_d) + \\
    + R^2(t) \dot{R}(t) \Bigg[ h_g(T_d) n_g(R(t),t) + \\ + \sum_{i=1}^k h_i(T_d) \left( n_i(R(t),t) - x_i v_l^{-1} (\{x\},T_d) \right) \Bigg].
\end{gathered}
\end{equation}
Recognizing equalities ${h_i = h_g}$, ${i=1,2\dots k}$, \eqref{total amount_n} and \eqref{dop1} allows us to rewrite eq.\eqref{dop3} as
\begin{equation}
\label{u_dop}
\begin{gathered}
   u(r,t)=\dot{R}(t) \frac{R^2(t)}{r^2} \Bigg[1+ \\ +\frac{1}{n_0 v_l(\{x\},T_d)} \frac{T_d}{T_0}\left( \frac{q(\{x\},T_d)}{h_g(T_d)} - 1 \right) \Bigg].
\end{gathered}
\end{equation}
As follows from eq.\eqref{u_dop} at ${\dot{R} > 0}$ with ${n_0 v_l \ll 1}$ and ${q(\{x\},T_d)/h_g(T_d) >1}$,
\begin{equation}
\label{sign}
    u(r,t) > 0,
\end{equation}
i.e., the velocity of the Stefan flow is directed from the growing droplet at any point outside the diffusion shell around the droplet.

Let us now consider the case when vapors are only small admixtures to the carrier gas, i.e., when the strong inequality
\begin{equation}
\label{promm1}
    \sum_{i=1}^k n_i(r,t) \ll n,
\end{equation}
fulfills, the relative deviations of the temperature of the vapor-gas medium due to release of the condensation heat are also small,
\begin{equation}
\label{promm2}
    | T(r,t) - T_0 | \ll T_0.
\end{equation}
Inequalities \eqref{promm1} and \eqref{promm2} together with eq.\eqref{promm3} and \eqref{vol} allow us to rewrite eq.\eqref{prom3} with a sufficient accuracy as
\begin{equation}
\label{dop4}
\begin{gathered}
     4 \pi h_g (T_0) n_{g0} r^2 u(r,t) \approx  - h_{l}(\{x\}, T_d) \dot{N} + \\ + 4 \pi R^2(t)\dot{R}(t) h_g (T_0) n_{g0} -\\
     - 4 \pi \sum_{i=1}^k \left( h_i (T_0) - h_g (T_0) \right) \int\limits_{R(t)}^{r} dr_1 r_1^2 \frac{\partial n_i(r,t)}{\partial t}.
\end{gathered}
\end{equation}
There is no material sources in the vapor-gas atmosphere, and we can write a continuity equation for each vapor,
\begin{equation}
\label{diff_flow}
    \frac{\partial n_i(r,t)}{\partial t} = -\nabla \vec{j}_i(r,t) = -\nabla \left[ \vec{J}_i(r,t) + n_i(r,t) \vec{u}(r,t) \right],
\end{equation}
where ${\vec{j}_i(r,t)}$ and ${\vec{J}_i(r,t)}$ are the densities of the total molecular flux and, respectively, the diffusion molecular flux of $i$th component of vapor. The integral term in eq.\eqref{dop4} can be rewritten with the help of eq.\eqref{diff_flow} as
\begin{equation}
\begin{gathered}
\label{dop5}
    4 \pi  \int\limits_{R(t)}^{r} dr_1 r_1^2 \frac{\partial n_i(r,t)}{\partial t} = 4 \pi \sum_{i=1}^k \left( j_i(R,t) R^2 - j_i(r,t)r^2 \right) = \\
    = 4 \pi \sum_{i=1}^k \left( j_i(R,t) R^2 - n_{i0}u(r,t)r^2 \right).
\end{gathered}
\end{equation}
Substituting eq.\eqref{dop5} in eq.\eqref{dop4}, recognizing smallness of the shift of the surface of droplet with the rate ${\dot{R}}$, using equalities
\begin{equation}
    \dot{N}_i = - 4 \pi R^2 \left( j_i(R,t) - n_i (R,t) \dot{R} \right) \approx - 4 \pi R^2 j_i(R,t),
\end{equation}
\begin{equation}
     4 \pi R^2 \sum_{i=1}^k j_i(R,t) = - \dot{N},
\end{equation}
eqs.\eqref{xx} and \eqref{q} yield
\begin{equation}
\label{prom4}
\begin{gathered}
  4 \pi h_g(T_0) n_{g0} r^2 u(r,t) \approx q(\{x\}, T_d) \dot{N} + \\
  + 4 \pi R^2(t) \dot{R}(t) h_g(T_0) n_{g0} - h_g (T_0) \dot{N},
\end{gathered}
\end{equation}
where we took into account that ${\sum\limits_{i=1}^k |h_i(T_0) - h_g(T_0)| n_i(r,t) \ll h_g(T_0) n_{g0} }$.
Replacing ${\dot{N}}$  by ${4 \pi R^2(t) \dot{R}(t) v_l^{-1}(\{x\}, T_d)}$, we find from \eqref{prom4}
{\sloppy}
\begin{equation}
\label{flow_rate-2}
\begin{gathered}
   u(r,t)=\dot{R}(t) \frac{R^2(t)}{r^2} \Bigg[1 + \\+ \frac{1}{n_{g0} v_l(\{x\},T_d)} \left( \frac{q(\{x\},T_d)}{h_g(T_0)}-1\right) \Bigg].
\end{gathered}
\end{equation}

According to procedure of derivation, eq.\eqref{flow_rate-2} is applicable outside the diffusion shell around the droplet where ${r \gg R(t)}$. Under conditions \eqref{promm1} and \eqref{promm2}, eq.\eqref{u_dop} reduces to eq.\eqref{flow_rate-2}. Thus eq.\eqref{sign} is still valid in the same domain, and we should observe the Stefan outflow with positive ${u(r,t)}$. Taking into account eq.\eqref{dop_6}, one can expect that velocity ${u(r,t)}$ should change its sign at some distance ${r_0}$ (${R(t) < r_0 }$) within the diffusion shell.
{\sloppy}

%%%%%%%%%%%%%%%%%%%%%%%%%%%%%%%%%%%%%%%%%%%%%%%%%%%%%%%%%%%%%%%%
%%%%%%%%%%%%%%%%%%%%%%%%%%%%%%%%%%%%%%%%%%%%%%%%%%%%%%%%%%%%%%%%

\section{A self-similar approach to the Stefan flow problem}\label{sec:self-similar}

\subsection{The local temperature and vapor concentrations in the vapor-gas mixture}\label{sec:local_temp}

To describe all the details in the behavior of the Stefan flow velocity ${u(r,t)}$ analytically at any distance from the droplet center, we will analyze the case of nonstationary self-similar regime of multicomponent nonisothermal droplet growth at fixed composition ${\{x\}}$ and temperature ${T_d}$. Let ${z}$ be the self-similar variable determined as
\begin{equation}
\label{z}
    z \equiv r/R(t),
\end{equation}
then self-similar vapor densities and temperature profile should be taken as
\begin{equation}
    n_i(r,t) = n_i(z)\quad (i=1,2\dots k), \quad T(r,t) = T(z).
\end{equation}
We will assume below that inequalities \eqref{promm1} and \eqref{promm2} fulfill. These inequalities can be rewritten in the self-similar variables in the form
\begin{equation}
\label{con1}
    \sum_{i=1}^k n_i(z) \ll n,
\end{equation}
\begin{equation}
\label{con2}
    \frac{T(z)-T_0}{T_0} \ll 1.
\end{equation}
Correspondingly, eq.\eqref{u_full_v} for the Stefan velocity profile transforms under conditions \eqref{con1}, \eqref{con2}, ${n_0v_l \ll 1}$ and self-similar variables as
\begin{equation}
\label{u_almost}
  u(z) \approx - \frac{\dot{R}(t)}{z^2}\left(
  \frac{1}{n_0v_l(\{x\},T_d)}+\frac{1}{T_0}\int\limits_{1}^{z} dz_1 z_1^3 \frac{dT(z_1,t)}{dz_1}\right).
\end{equation}

To find the velocity ${u(r,t)}$ with the help of eq.\eqref{u_almost}, it is sufficient to derive expressions for rate ${\dot{R}(t)}$ and non-stationary temperature profile ${T(z)}$  in the vapor-gas atmosphere  in the lowest approximation with respect the small parameter ${n_0v_l \ll 1}$. In this case, we can use the equations of diffusion and heat transfer obtained with neglecting the Stefan flow. Then nonstationary equations for the diffusion to the droplet and the heat transfer from the droplet to the medium can be written in the following form (\cite{Grinin2008a, Grinin2011a, Kuchma2015a})
\begin{equation}
\label{dn}
\begin{gathered}
    \frac{\partial n_i(r,t)}{\partial t} = \frac{D_i}{r^2}\frac{\partial}{\partial r}\left[r^2 \frac{\partial n_i(r,t)}{\partial r}\right] - \\ - \frac{\dot{R}R^2(t)}{r^2}\frac{\partial n_i(r,t)}{\partial r},
    \quad i=1,2\dots k,
\end{gathered}
\end{equation}
\begin{equation}
\label{dTT}
    \frac{\partial T(r,t)}{\partial t} = \frac{\chi}{r^2}\frac{\partial}{\partial r}\left[r^2 \frac{\partial T(r,t)}{\partial r}\right] - \frac{\dot{R}R^2(t)}{r^2}\frac{\partial T(r,t)}{\partial r},
\end{equation}
where ${D_i}$ is the diffusivity of ${i}$th component and ${\chi}$ is the thermal diffusivity of the vapor-gas medium.
The boundary conditions at the droplet surface at ${r=R}$ for eqs.\eqref{dn} and \eqref{dTT} are the equations of material and heat balance at the droplet surface
\begin{equation}
\label{bond_1}
\begin{gathered}
    \dot{N} = \frac{d}{dt}\left(\frac{4 \pi R^3}{3 v_l(\{x\},T_d)}\right) = \sum_{i=1}^k \dot{N}_i = \\ =\left(4 \pi r^2 \sum_{i=1}^k D_i \frac{\partial n_i(r,t)}{\partial r}\right)\Bigg |_{r=R},
\end{gathered}
\end{equation}
\begin{equation}
\label{bond_2}
    \left(\kappa \frac{\partial T(r,t)}{\partial r}\right)\Bigg |_{r=R} = - \left( \sum_{i=1}^k q_{il}(\{x\},T_d) D_i \frac{\partial n_i(r,t)}{\partial r}\right)\Bigg |_{r=R}.
\end{equation}
Here ${\kappa}$ is the thermal conductivity of the vapor-gas mixture.
Equations \eqref{dn} and \eqref{dTT} and the boundary conditions \eqref{bond_1}, \eqref{bond_2} rewritten in the self-similar variables have the form
\begin{equation}
\label{dnzz}
    \frac{d^2 n_i(z)}{d z^2} + \left[\frac{2}{z} + \frac{R\dot{R}}{D_i}\left(z - \frac{1}{z^2}\right)\right]\frac{dn_i(z)}{dz} = 0, \quad i=1,2\dots k,
\end{equation}
\begin{equation}
\label{dTzz}
    \frac{d^2 T(z)}{d z^2} + \left[\frac{2}{z} + \frac{R\dot{R}}{\chi}\left(z - \frac{1}{z^2}\right)\right]\frac{dT(z)}{dz} = 0
\end{equation}
and
\begin{equation}
\label{bond1}
    R\dot{R} = v_l(\{x\},T_d) \sum_{i=1}^k D_i \frac{d n_i(z)}{d z}\Bigg |_{z=1},
\end{equation}
\begin{equation}
\label{bond_2.1}
    \kappa \frac{d T(z)}{d z}\Bigg |_{z=1} = -  \sum_{i=1}^k q_{il}(\{x\},T_d) D_i \frac{d n_i(z)}{dz}\Bigg |_{z=1}.
\end{equation}
Note, as follows from eq.\eqref{bond1}, the rate ${R\dot{R}}$ (entering eqs.\eqref{dnzz} and \eqref{dTzz})  is independent of ${z}$, but generally depends on droplet composition and temperature. This fact is important for existence of the self-similar solution. It allows simultaneous solving of ordinary differential equations  eq.\eqref{dnzz} and eq.\eqref{dTzz} together with conditions \eqref{cond_ni} and \eqref{cond_T}. As a result we have
\begin{equation}
\label{n_ss}
\begin{gathered}
     n_i(z)=n_{i\infty}(\{x\},T_d)+(n_{i0}-n_{i\infty}(\{x\},T_d)) \times \\
     \times \frac{\displaystyle\int\limits_1^z \frac{dy}{y^2}exp\left[-\frac{R\dot{R}}{2D_i}     \left(y^2+\frac{2}{y}-3\right)\right]}{\displaystyle\int\limits_1^{\infty} \frac{dy}{y^2}exp\left[-\frac{R\dot{R}}{2D_i}\left(y^2+\frac{2}{y}-3\right)\right]},
     \quad i=1,2 \dots k,
\end{gathered}
\end{equation}
\begin{equation}
\label{T(z)}
     T(z)=T_d+(T_0-T_d)\frac{\displaystyle\int\limits_1^z \frac{dy}{y^2}exp\left[-\frac{R\dot{R}}{2\chi}     \left(y^2+\frac{2}{y}-3\right)\right]}{\displaystyle\int\limits_1^{\infty} \frac{dy}{y^2}exp\left[-\frac{R\dot{R}}{2\chi}\left(y^2+\frac{2}{y}-3 \right) \right]}.
\end{equation}
Here we took into account the conditions \eqref{cond_ni} and  \eqref{cond_T}.

%%%%%%%%%%%%%%%%%%%%%%%%%%%%%%%%%%%%%%%%%%%%%%%%%%%%%%%%%%%%%%%%
%%%%%%%%%%%%%%%%%%%%%%%%%%%%%%%%%%%%%%%%%%%%%%%%%%%%%%%%%%%%%%%%

\subsection{Stationary composition and temperature of the droplet}\label{sec:stat}

The stationary rate ${R\dot{R}}$ of droplet growth and the droplet temperature ${T_d}$ satisfy two coupled integral equations following from eqs.\eqref{bond1} and \eqref{bond_2.1}
\begin{equation}
\label{statR}
    R\dot{R} = v_l(\{x\} \sum_{i=1}^k \frac{D_i (n_{i0}-n_{i\infty}(\{x\},T_d))}{\displaystyle\int\limits_1^{\infty} \frac{dy}{y^2}exp\left[-\frac{R\dot{R}}{2D_i}\left(y^2+\frac{2}{y}-3 \right) \right]},
\end{equation}
\begin{equation}
\label{statT}
\begin{gathered}
     T_d - T_0= - \displaystyle\int\limits_1^{\infty} \frac{dy}{y^2}exp\left[-\frac{R\dot{R}}{2D_i}\left(y^2+\frac{2}{y}-3 \right) \right] \times \\
     \times \sum_{i=1}^k \frac{q_{il}(\{x\},T_d) D_i (n_{i0}-n_{i\infty}(\{x\},T_d))}{\kappa \displaystyle\int\limits_1^{\infty} \frac{dy}{y^2}exp\left[-\frac{R\dot{R}}{2D_i}\left(y^2+\frac{2}{y}-3 \right) \right]}.
\end{gathered}
\end{equation}

Finally we need an equation for the stationary droplet composition which is established independently of initial composition of the growing droplet. As follows from definition \eqref{xx}, the condition of a fixed composition in the droplet can be written as
\begin{equation}
\label{dx}
   \dot{x}_i = \frac{\dot{N}_i(t) - x_i\dot{N}(t)}{N(t)} = 0, \quad i =1, 2 \dots k,
\end{equation}
In view of eqs.\eqref{bond_1} and \eqref{n_ss},
\begin{equation}
\label{dN}
\begin{gathered}
   \dot{N}_i(t) = 4 \pi R \frac{D_i (n_{i0}-n_{i\infty}(\{x\},T_d))}{\displaystyle\int\limits_1^{\infty} \frac{dy}{y^2}exp\left[-\frac{R\dot{R}}{2D_i}\left(y^2+\frac{2}{y}-3 \right) \right]}, \\ i =1, 2 \dots k.
\end{gathered}
\end{equation}
Taking into account eqs.\eqref{bond_1} and \eqref{dN}, we find from eq.\eqref{dx}  ${k}$ equations for the composition in the growing droplet
\begin{equation}
\label{xs}
\begin{gathered}
    \frac{1}{x_i}\frac{D_i \left(n_{i0}-n_{i\infty}(\{x\},T_d)\right)}{\displaystyle\int\limits_1^{\infty} \frac{dy}{y^2}exp\left[-\frac{R\dot{R}}{2D_i}\left(y^2+\frac{2}{y}-3 \right) \right]} = \\
    = \sum_{m=1}^k \frac{D_m \left(n_{m0}-n_{m\infty}(\{x\},T_d)\right)}{\displaystyle\int\limits_1^{\infty} \frac{dy}{y^2}exp\left[-\frac{R\dot{R}}{2D_m}\left(y^2+\frac{2}{y}-3 \right) \right]}, \quad
    i=1,2\dots k.
\end{gathered}
\end{equation}
If we know the explicit thermodynamical expressions for ${v_l(\{x\},T_d)}$, ${n_{i\infty}(\{x\},T_d)}$ and ${q_{il}(\{x\},T_d)}$ at ${i=1,2\dots k}$, eqs. \eqref{statR}, \eqref{statT}, \eqref{xs} together with eqs.\eqref{u_almost} and \eqref{T(z)} form a complete set of equations for the problem of finding the profile of the Stefan flow velocity around a growing multicomponent droplet.

To simplify the Stefan velocity profile problem at the characteristic values of the parameters in the case of growing droplet in the vapor-gas atmosphere, the following approximations can be used,
\begin{equation}
\label{appr}
\begin{gathered}
   \int\limits_1^{\infty} \frac{dy}{y^2}exp\left[-\frac{R\dot{R}}{2\chi}\left(y^2+\frac{2}{y}-3\right)\right]\approx 1,\\
    \int\limits_1^{\infty} \frac{dy}{y^2}exp\left[-\frac{R\dot{R}}{2 D_i}\left(y^2+\frac{2}{y}-3\right)\right] \approx 1.
\end{gathered}
\end{equation}
Substituting eq.\eqref{appr} into eq.\eqref{T(z)} for the local temperature profile in a vapor-gas mixture gives
\begin{equation}
\label{Tz}
     T(z) \approx T_d-(T_d-T_0)\int\limits_1^z \frac{dy}{y^2}exp     \left(-\frac{R\dot{R}}{2\chi}y^2\right),
\end{equation}
where the stationary droplet growth rate ${R\dot{R}}$ satisfies according to eq.\eqref{statR} to the formula
\begin{equation}
\label{RR}
     R\dot{R} = v_l(\{x\},T_d)\sum_{i=1}^k D_i \left( n_{i0} - n_{i\infty}(\{x\},T_d) \right).
\end{equation}

Substituting eq.\eqref{appr} into eqs.\eqref{statT} and \eqref{xs} for the stationary droplet temperature ${T_d}$ and droplet composition ${\{x\}}$ gives
\begin{equation}
\label{stat_1}
   \kappa (T_d-T_0) = \sum\limits_{i=1}^k q_{il} D_i \left( n_{i0} - n_{i\infty}(\{x\},T_d) \right),
\end{equation}
\begin{equation}
\label{stat_2}
 \begin{gathered}
    \frac{D_i \left( n_{i0} - n_{i\infty}(\{x\},T_d)\right)}{x_i} = \frac{D_1 \left( n_{10} - n_{1\infty}(\{x\},T_d) \right)}{x_1},\\ \quad    i = 1,2 \dots k, \\ \sum\limits_{i=1}^k x_i =1.
 \end{gathered}
\end{equation}
%Hence, the combination of eq.\eqref{RR}-\eqref{stat_2} constitutes a system for description of the growth of suspended spherical droplet by diffusion, which can be solved knowing initial conditions ${T_0}$, ${n_{i0}}$, ${\{x_0\}}$, ${R_0}$.

%%%%%%%%%%%%%%%%%%%%%%%%%%%%%%%%%%%%%%%%%%%%%%%%%%
\subsection{The Stefan velocity profile}\label{sec:velocity}
%%%%%%%%%%%%%%%%%%%%%%%%%%%%%%%%%%%%%%%%%%%%%%%%%%%5

Differentiating eq.\eqref{Tz} with respect to variable ${z}$ yields
\begin{equation}
\label{dTz}
   \frac{d T(z_1)}{dz_1} \approx -\frac{T_d-T_0}{z_1^2}exp \left(-\frac{R\dot{R}}{2\chi}z_1^2\right).
\end{equation}
Substituting eqs.\eqref{dTz} into eq.\eqref{u_almost} for the velocity of the Stefan flow ${u(r,t)}$ and using equality ${h_g(T_0) = c_g T_0}$, eqs.\eqref{z} and \eqref{stat_1} allow one to obtain
\begin{equation}
\label{uzl}
\begin{gathered}
  u(z) \approx  \frac{\dot{R(t)}}{z^2}\frac{1}{v_{l}(\{x\},T_d)
  n_0}\Bigg[\frac{q(\{x\},T_d)}{c_gT_0} \times \\ \times \left(1- exp \left(-\frac{R\dot{R}}{2\chi}z^2\right)\right)-1\Bigg].
\end{gathered}
\end{equation}

The distance ${r_{u=0}}$ from the center of the droplet, where the Stefan flow velocity changes its sign, corresponds to the value ${z = z_0}$ at which ${u(z_0)=0}$. As follows from eqs.\eqref{z} and \eqref{uzl}
\begin{equation}
\label{r_0}
  \frac{r_{u=0}^2}{R^2(t)} =- \frac{2\chi}{R\dot{R}}\ln\left(1-\frac{c_gT_0}{q(\{x\},T_d)}\right).
\end{equation}

Outside the non-stationary diffusion shell around the droplet, where ${\displaystyle \frac{R\dot{R}}{2\chi}z^2 \gg 1}$, eq.\eqref{uzl} for the Stefan flow velocity reduces to
\begin{equation}
  u(z) \approx  \frac{\dot{R(t)}}{z^2}\frac{1}{v_{l}(\{x\},T_d)
  n_0}\left(\frac{q(\{x\},T_d)}{c_gT_0}-1\right),
\end{equation}
or, in spherical coordinates, to
\begin{equation}
\label{uuu}
  u(r,t) \approx \dot{R(t)} \frac{R^2(t)}{r^2}\frac{1}{v_{l}(\{x\},T_d)
  n_0}\left(\frac{q(\{x\},T_d)}{c_gT_0}-1\right).
\end{equation}
Eq.\eqref{uuu} coincides with a general asymptotic form given by eq.\eqref{flow_rate-2} at ${T_d-T_0 \ll T_0}$.
{\sloppy

}
In the vicinity of the droplet, where ${\displaystyle \frac{R\dot{R}}{2\chi}z^2 \ll 1}$, eq.\eqref{uzl} reduces to
\begin{equation}
  u(r,t) \approx  - \frac{\dot{R} }{n_0 v_{l}(\{x\},T_d)},
\end{equation}
which coincides with another general asymptotic form given by eq.\eqref{dop_6} at ${T_d-T_0 \ll T_0}$.
%We want to point out that Stefan's flow near the  droplet is directed to the droplet, which provide motion of small aerosol particles to the growing droplet, and, from the other hand, Stefan's flow far from droplet is directed away as consequence of the condensation heat release, which lead to expansion of the medium around the growing droplet.\

Let us consider now the expansion of the vapor-gas medium around the growing droplet. The vapor-gas mixture expands in response to Stefan's outflow. The corresponding increase of the volume of the vapor-gas mixture per a droplet ${\Delta V(t)}$ at a given velocity profile ${u(r,t)}$ (outside the non-stationary diffusion shell) can be estimated as
\begin{equation}
\label{volume_add}
   \Delta V(t)=4\pi  \int\limits_{0}^{t}r^2 u(r,t_1)dt_1.
\end{equation}
Substituting eq.\eqref{uuu} into eq.\eqref{volume_add} and using eq.\eqref{vol}, we obtain for the volume increase ${\Delta V(t)}$
\begin{equation}
\label{add_volume}
   \Delta V(t)\approx \frac{1}{n_0 v_l(\{x\},T_d)} \left( \frac{q(\{x\},T_d)}{c_gT_0}-1\right) V_d(t).
\end{equation}\\

The presented calculation of the hydrodynamic flow of the medium illustrates in detail the mechanism of increasing the volume of vapor-gas system at non-isothermal condensational growth of a single droplet. Let us now note that the idea of the independence of individual droplets in the process of condensation of supersaturated vapor is true, strictly speaking, only on the initial part of the stage of nucleation of the droplets.  By the completion of the nucleation stage, a significant overlapping of diffusion layers of separate droplets begins. As shown by \cite{Kuchma2015a}, in the case of vapor condensation markedly below the critical point, when the role of non-stationary diffusion of vapors in the dynamics of droplet growth is small, the influence of overlapping of diffusional shells can be taken into account within the mean-field approximation of vapor supersaturation and temperature. Concentrations of vapors and temperature of the vapor-gas medium are assumed in this approximation to be homogeneous in space. In this case, the evaluation of the thermal expansion of the medium can be derived directly from the conservation laws. If the number of condensed molecules ${N}$ is small compared with their total number ${n_0 V_0}$ in the initial volume ${V_0}$ in the vapor-gas mixture, the volume change of vapor-gas medium ${\Delta V}$, in the approximation of its ideality and at constant pressure, can be written as
\begin{equation}
   \frac{\Delta V(t)}{V_0} = -\frac{N}{n_0 V_0} + \frac{\Delta T}{T_0},
\end{equation}
where ${\Delta T \ll T_0}$ is the corresponding change of the medium temperature. The value ${\Delta T}$ is determined, in turn, by the balance condition of the heat released during condensation, that has, at fixed composition and temperature of droplets, a form
\begin{equation}
   q (\{x\}, T_d) N \approx c_g n_0 V_0 \Delta T.
\end{equation}
As a result, we find
\begin{equation}
\begin{gathered}
   \Delta V \approx \left( \frac{q(\{x\},T_d)}{c_gT_0}-1\right) \frac{N}{n_0} = \\ = \left( \frac{q(\{x\},T_d)}{c_gT_0}-1\right) \frac{1}{n_0 v_l(\{x\},T_d)} V_l,
\end{gathered}
\end{equation}
where ${V_l}$ is the total volume of condensed droplets. The resulting expression is equivalent to eq.\eqref{add_volume} in the case of a single drop, and it may give the impression that for description of the thermal expansion of the cloud in general is not necessary to consider the behavior of individual droplets. Note, however, that we used in the assessment the values ${q (\{x\}, T_d)}$  and ${v_l (\{x\}, T_d)}$ which depend on composition and temperature of droplets (at this, the dependence on the composition of the droplet is most significant). These values can be determined, as shown above in this section, only by corresponding equations of evolution for a single droplet.
{\sloppy}

%%%%%%%%%%%%%%%%%%%%%%%%%%%%%%%%%%%%%%%%%%%%%%%%%%%%%%%%%%%%%%%%
%%%%%%%%%%%%%%%%%%%%%%%%%%%%%%%%%%%%%%%%%%%%%%%%%%%%%%%%%%%%%%%%

\section{Numerical illustration of the approach for specific binary mixtures of vapors in the air }\label{sec:simulating}

Let us illustrate our approach to the problem of the Stefan multicomponent outflow in the case of droplet growth in the atmosphere of air and mixtures of several vapors. We will characterize the presence of vapor by its supersaturation ${\zeta_i}$ with respect to the equilibrium at flat interface of its pure liquid at temperature ${T_0}$
\begin{equation}
\label{saturation}
    \zeta_i \equiv \frac{n_{i0}-n_{i\infty}(x_i=1,T_0)}{n_{i\infty}(x_i=1,T_0)}.
\end{equation}
According to eqs.\eqref{statR}--\eqref{xs}, assignment of temperature and vapor saturations in the vapor-gas atmosphere determines uniquely the stationary droplet growth rate ${R\dot{R}}$, the droplet temperature ${T_d}$ and composition ${\{x\}}$. As a next step, using eqs.\eqref{uzl}, \eqref{r_0} and \eqref{add_volume} allows us to find evolution in time of the Stefan flow velocity ${u(r,t)}$, the distance ${r_{u=0}}$ and volume increase ${\Delta V(t)}$. Below we will compute these quantities for a binary droplet in air with binary mixture of ethanol-water (${C_2H_5OH/H_2O}$) vapors, binary mixture of sulfuric acid-water (${H_2SO_4/H_2O}$) vapors, and with pure water vapor (${H_2O}$).
{\sloppy}

%We will compute these quantities for systems
%\begin{enumerate}
%   \item Air and two-component vapor (ethanol ${i=1}$, water ${i=2}$), where ${\zeta_1=\zeta_2=1}$, i.e. vapors are supersaturated.
%   \item Air and two-component vapor (ethanol ${i=1}$, water ${i=2}$), where ${\zeta_1=-0.5}$, ${\zeta_2=-0.3}$, i.e. vapors are subsaturated.
%   \item Air and two-component vapor (sulfuric acid ${i=1}$, water ${i=2}$), where ${\zeta_1=10}$, ${\zeta_2=-0.3}$, i.e. sulfuric acid vapor is supersaturated and water vapor is undersaturated.
%   \item Air and supersaturated water vapor ${\zeta=1}$.
%\end{enumerate}

The results of our computations are presented in Tab.~\ref{tab:table1}. Let us recall that we consider the stationary droplet growth after small transient stage, with the stationary droplet composition ${\{x\}}$ and temperature ${T_d}$ which do not depend on the droplet size. Specific thermodynamic data for the considered substances and their ideal mixtures in gaseous and liquid states were taken from \cite{Taleb1996, Jaecker1990, Perry2008} ((see the details in \cite{Martyukova2016a}). The dynamics of stationary-state establishment can be complicated and can differ for various systems (\cite{Martyukova2016a}), however, the Stefan flow velocity and, consequently, the additional volume have been derived assuming a stationary temperature and composition of the droplet established after the transient stage.

\begin{widetext}
\begin{table}[h]

\caption{\label{tab:table1} The values of stationary composition ${x_1}$, temperature ${T_d}$ and growth rate of the droplet ${R\dot{R}}$, the distance ${r_{u=0}(t)/R(t)}$ and the relative volume increase ${\Delta V(t)/V_d(t)}$ in air with vapor mixtures ${C_2H_5OH/H_2O}$, ${H_2SO_4/H_2O}$, ${H_2O}$ at the same bulk temperature ${T_0=293}$K and vapor-air pressure ${P = 1}$ atm.}

\begin{tabularx}{\textwidth}{ p{2cm} ||X|X|X|X}
\hline
${i=1}$ & ${C_2H_5OH }$ & ${C_2H_5OH}$ & ${H_2SO_4}$ & ${H_2O}$  \\
${i=2}$ & ${H_2O}$ & ${H_2O}$ & ${H_2O}$ & ${-}$\\
 \hline
 \hline
 $ \mathbf{\zeta_1} $ & 1 & -0.5 & 10 & 1 \\
 $ \mathbf{\zeta_2} $ & 1 & -0.3 & -0.3 & - \\ \hline
% $ \boldsymbol{R_0} $, cm & $5 \cdot 10^{-4}$ & $5 \cdot 10^{-4}$ & $10^{-3}$ & $5 \cdot 10^{-4}$ \\
$ \boldsymbol{x_{1}} $ & 0.51 & 0.42 & 0.3 & 1 \\
$ \boldsymbol{T_{d}} $, K & 313 & 295 & 293 & 301 \\
$ \boldsymbol{R\dot{R}}$, & $4.7 \cdot 10^{-6}$ & $4.5 \cdot 10^{-7}$ & $3.8 \cdot 10^{-11}$ & $10^{-6}$ \\
$ {\mathrm{cm^2s^{-1}}}$ &  &  &  &  \\
$ \boldsymbol{r_{u=0}(t)/R(t)} $ & 136 & 438 & $4.2 \cdot 10^{4}$ & 151 \\
$ \boldsymbol{\Delta V(t)/V_d(t)} $ & $2.5 \cdot 10^{3}$ & $2.9 \cdot 10^{3}$ & $4.7 \cdot 10^{3}$ & $5.6 \cdot 10^{3}$
\end{tabularx}

\end{table}
\end{widetext}
As we can see, the results are really related to a growing droplet, because ${R\dot{R} > 0}$ for all systems considered in Tab.~\ref{tab:table1}, even in the case when both vapors are undersaturated (${C_2H_5OH/H_2O}$ (2)). In view of huge difference in saturated vapor pressure between sulfuric acid vapor and water, ${P_{H_2SO_4}/P_{H_20} \sim 10^{-6}}$, the stationary rate of droplet growth in the mixture ${H_2SO_4/H_20}$ is much lower than for other droplets (${C_2H_5OH/H_2O}$, ${H_2O}$).
{\sloppy}

As seen from Tab.~\ref{tab:table1}, the distance ${r_{u=0}}$, where the Stefan flow changes its direction, increases with time proportionally to the droplet radius ${R(t)}$. For lower droplet growth rate the distance ${r_{u=0}}$ is larger than for others what is evident directly from eq.\eqref{r_0}. Also as consequence of droplet growth the additional volume increase ${\Delta V(t)}$ grows in time proportionally to the volume of the droplet. We note that the maximal coefficients ${\Delta V(t)/V(t)}$ are reached for  ${C_2H_5OH/H_2O}$ (1) and ${H_2O}$ droplets.
{\sloppy}
\begin{figure}[h]
\begin{center}
    \includegraphics[width=0.96\linewidth]{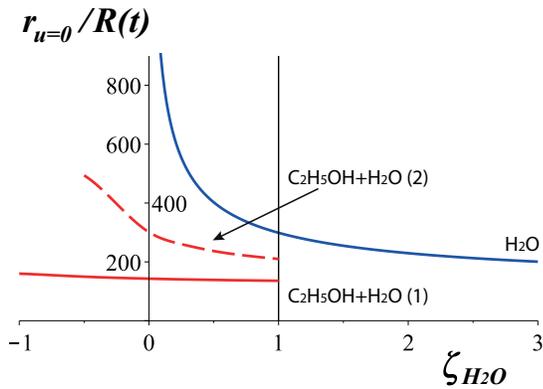}
    \caption{\label{fig:r_0} The scaled distance ${r_{u=0}(t)/R(t)}$ as a function of water vapor supersaturation ${H_2O}$ droplet and binary  ${C_2H_5OH/H_2O}$ droplet at ${\zeta_{C_2H_5OH} = 1}$ (1), ${\zeta_{C_2H_5OH} = -0.5}$ (2).}
\end{center}
\end{figure}

Fig.~\ref{fig:r_0} shows how the distance ${r_{u=0}(t)/R(t)}$ depends on water vapor supersaturation.  The pure water droplet grows if water vapor supersaturation is more than ${0}$ (blue line in Fig.~\ref{fig:r_0}). The binary ethanol-water droplet can grow depending on the ethanol vapor supersaturation. If ${\zeta_{C_2H_5OH} = 1}$ (case 1, solid red line in Fig.~\ref{fig:r_0}), the droplet grows at any water supersaturation ${\zeta_{H_20} > 1}$, if ${\zeta_{C_2H_5OH} = -0.5}$ (case 2, dash red curve in Fig.~\ref{fig:r_0}), the droplet grows at ${\zeta_{H_20}> -0.5}$. The water vapor flux to the droplet increases with an increase of water vapor supersaturation for any droplet, therefore, the quantity ${r_{u=0}/R(t)}$ becomes smaller, and curves in Fig.~\ref{fig:r_0} decrease monotonically with an increase of ${\zeta_{H_2O}}$ .{\sloppy}

\section{Conclusions}\label{sec:conclusions}

We have presented in this paper a theoretical study of the Stefan flow caused by a free growing droplet in the vapor-gas atmosphere with several condensing components. Using the nonstationary heat and material balance and diffusion transport equations, we have derived an analytical expression for multicomponent Stefan's flow and have established that the Stefan flow goes to the growing droplet in the droplet vicinity, and transforms to outflow at some distance ${r_{u=0}}$ from the droplet as a consequence of nonstationary heat conduction of the heat releasing at vapor condensation in the droplet. For an ensemble of droplets in the atmospheric cloud, such effect provides an increase of the total volume of the cloud, which can be treated as cloud thermal expansion and leads to to rise of the cloud as a whole due to buoyancy.\\

\section*{Acknowledgement}

This work was supported by St. Petersburg State University under grant 11.37.183.2014.

%%%%%%%%%%%%%%%%%%%%%%%%%%%%%%%%%%%%%%%%%%%%%%%%%%%%%%%%%%%%%%%%
%%%%%%%%%%%%%%%%%%%%%%%%%%%%%%%%%%%%%%%%%%%%%%%%%%%%%%%%%%%%%%%%

\bibliographystyle{elsarticle-harv}

\end{document}